\documentclass[10pt,preprintnumbers,amsmath,amssymb,showpacs,prb,twocolumn]{revtex4}
\usepackage{stmaryrd}
\usepackage{amsmath}
\usepackage{amssymb}
\usepackage{float}
\usepackage{array, color}
\usepackage{tabularx}
\usepackage{threeparttable}
\usepackage{titletoc}
\usepackage{booktabs}
\usepackage[colorlinks,linkcolor=blue,anchorcolor=blue,citecolor=blue,urlcolor=blue,dvipdfm]{hyperref}

\usepackage{graphicx}
\usepackage{subfigure}
\begin{document}

\title{Adsorption of molecular oxygen on doped graphene: atomic, electronic and magnetic properties}
\author{Jiayu Dai and Jianmin Yuan}
\email{jmyuan@nudt.edu.cn}
 \affiliation{ Department of Physics, National University
of Defense Technology, Changsha 410073, People's Republic of China}

\date{November 24, 2009}
\begin{abstract}

Adsorption of molecular oxygen on B-, N-, Al-, Si-, P-, Cr- and
Mn-doped graphene is theoretically studied using density functional
theory in order to clarify if O$_2$ can change the possibility of
using doped graphene for gas sensors, electronic and spintronic
devices. O$_2$ is physisorbed on B-, and N-doped graphene with small
adsorption energy and long distance from the graphene plane,
indicating the oxidation will not happen; chemisorption is observed
on Al-, Si-, P-, Cr- and Mn-doped graphene. The local curvature
caused by the large bond length of X-C (X represents the dopants)
relative to C-C bond plays a very important role in this
chemisorption. The chemisorption of O$_2$ induces dramatic changes
of electronic structures and localized spin polarization of doped
graphene, and in particular, chemisorption of O$_2$ on Cr-doped
graphene is antiferromagnetic. The analysis of electronic density of
states shows the contribution of the hybridization between O and
dopants is mainly from the $p$ or $d$ orbitals. Furthermore, spin
density shows that the magnetization locates mainly around the doped
atoms, which may be responsible for the Kondo effect. These special
properties supply a good choice to control the electronic properties
and spin polarization in the field of graphene engineering.
\end{abstract}

 \maketitle

\section{Introduction}

Graphene has been becoming a new star in a lot of fields after its
successful fabrication\cite{first,Rise,Neto}, especially for the
application of two aspects below because of the many reasons for the
renewed interest: first, it is very potential to apply graphene as
gas sensors with high sensitivity because the transport properties
exhibit large changes upon exposure to several gases such as
NO$_2$\cite{Schedin}. Graphene can be used as an excellent sensor
material because of its special properties such as two dimensional
property maximizing the interaction of adsorbates on the layer, low
Johnson noise\cite{Rise,Novoselov,Zhang,Danneau}, and few crystal
defects\cite{Rise,Novoselov,Zhang}. This tells us the electronic
properties of graphene system can be sensitive to the adsorption of
gases. Second, there is an extraordinary interest in the electronic
band properties and magnetic order in graphene-related materials
that can be put into use as the next-generation electronic devices
and recording media, magnetic inks, spin qubits and spintronic
devices\cite{Neto,control,spin}, which can use the advantages of
high mobility of electrons in graphene and long coherence times in
carbon-based materials. A few investigations have been performed
until now to search for the potential application of graphene or
graphene nanoribbons as gas sensors\cite{sensor,Leenanerts}.
However, it has been shown that intrinsic graphene can only
physisorb most of gas molecules\cite{Leenanerts,Paolo,OurPaper} and
has no band gap and spin polarization. This, for example, prevents
the use of graphene in making gas sensors, transistors and
spintronic devices.

In order to improve the sensitivity for gases and the electronic
structures in graphene and carbon nanotubes (CNTs), the method of
doping is usually used. B- and N-doped graphene can also improve gas
sensing of CO, NO, NO$_2$ and NH$_3$\cite{LDA}; Al-doped graphene
can be sensitive to most of common gases in air\cite{apl}; H$_2$O on
Ti-doped graphene\cite{Rangel}, chlorophenols on Si-doped
CNTs\cite{si} are also reported. Meanwhile, the electronic
structures and magnetic properties in graphene and graphene
nanoribbons can be changed by doping different atoms such as
boron\cite{half-B}, sulfur and phosphorus\cite{Garcia}, transition
metal atoms\cite{tran,rev-tran,zhong}, or by defect
engineering\cite{defect}, functionalized with different atoms,
molecules and clusters\cite{func} and so on. Experimentally, several
CNTs and graphene-based materials with different dopants such as
nitrogen\cite{Nitr}, boron\cite{bn}, phosphorus\cite{pho} were also
synthesized. It has been shown by these studies that the doped
graphene or CNTs materials are very potential for gas sensors,
electronics and spintronics.

However, in order to use these materials in reality, the effect of
O$_2$ should be considered, since O$_2$ is one of the most important
gases taking up more than 20\% in air. Meanwhile, it has been shown
that O$_2$ can hole-dope semiconductors\cite{hole}. Also, gas
molecules can change the electronic and magnetic properties of
graphene and graphene nanoribbons\cite{gas}. Previous
calculations\cite{Paolo} and experiments\cite{exp} have shown that
O$_2$ in the triplet state is physisorbed on the surface of
single-walled carbon nanotubes (SWCNTs) and graphene, and the
transport properties could not be significantly affected. On the
other hand, the electronic properties of carbon-doped boron nitride
nanotubes can be changed dramatically by the chemisorption of O$_2$
molecule\cite{cbn}. This indicates that the dopants can improve the
reactivity of materials, and the O$_2$ molecules can give rise to
very different properties. Therefore, how O$_2$ molecule can affect
the properties of doped graphene is necessary to be understood in
the fields of both gas sensing and electronics.

In this paper we study from first principles the adsorption of
molecular oxygen on B-, N-, Al-, Si-, P-, Cr- and Mn-doped graphene.
In order to illuminate easily, these materials are abbreviated to be
BG, NG, AG, SG, PG, CG and MG. BG and NG retain a planar form, while
other atoms protrude out of the graphene layer and induce a local
curvature in graphene. Furthermore, O$_2$ molecule is physisorbed on
BG and NG with relatively small adsorption energy and large distance
of X-O (X represents the doping atom), comparing with chemisorption
on the doped graphene with other dopants. In the end, the
chemisorption induces dramatic change for the electronic structures
of doped graphene, and injects magnetic moments into the system
except O$_2$ on PG. Especially, the system of O$_2$ on CG is
antiferromagnetic (AFM).

\section{Computational Methods}

Spin-polarized density-functional theory (DFT) calculations are
performed using the Perdew-Burke-Ernzerhof (PBE)\cite{PBE}
generalized gradient approximation (GGA) for the
exchange-correlation potential. A supercell of $6\times6$ graphene
including 72 C atoms with a doped atom substituting a C atom and a
single O$_2$ molecule adsorbed onto it is constructed. With this
model, the dopant concentration in our calculations is $\sim 1.4$\%.
Experiments have shown the existence of these types of single-atom
doped materials\cite{single}. Besides, a method to synthesize the
transition metals doped graphene has been proposed\cite{rev-tran}.
In the direction normal to the surface, 29.5 {\AA} length in the
supercell is sufficient to minimize the interaction between graphene
layers. Ultrasoft pseudopotentials\cite{pot} and a plane-wave basis
set-up to a kinetic energy cutoff of 25 Ry for the wave function and
of 200 Ry for the charge density are chosen in all simulations (for
transition metals Cr and Mn, a cutoff of 35 Ry for the wave function
and 300 Ry for the charge density). The Brillouin zone is sampled
using a $3\times3\times1$ Monkhorst-Pack\cite{MK} grid and
Methfessel-Paxton\cite{MP} smearing of 0.01 Ry. A denser
$11\times11\times1$ Monkhorst-Pack grid and the tetrahedron
method\cite{tetrahedron} are used for the calculation of density of
states (DOS) and partial DOS (PDOS). Atomic positions are optimized
until the maximum force on any atom is less than 0.001 a.u. All
calculations are performed using the Quantum-ESPRESSO package
\cite{pwscf}. We carefully test this supercell for the convergence
of energy, magnetic properties, comparing a larger supercell as in
Ref.~\onlinecite{rev-tran}, and there is few difference found.

Here, we would like to point out that the usage of GGA, the
consequent neglect of van der Waals interactions, leads to an
incorrect description of physisorption, but this is of little
concern for us since we are interested in chemically bound
molecules. In fact, in physisorbed systems, LDA results look
``better'' than GGA, but they are just ``wrong in a different way''.
For chemically bound systems, GGA is usually a better choice.

In this calculation, only the triplet O$_2$ molecule is considered,
which is the ground state of O$_2$. The adsorption energy is defined
by $E_a=E_{tot}-E_{dg}-E_o$, where $E_{tot}$ is the total energy of
the doped graphene with a bound O$_2$ molecule, $E_{dg}$ is the
energy of doped graphene and $E_o$ is the energy of isolated O$_2$
molecule. In order to minimize systematic errors, the same
supercells and $k$-point grids are used for all calculations.

\section{Results and discussion}

\begin{figure}[!tb]
\centering
\includegraphics*[width=3.5in]{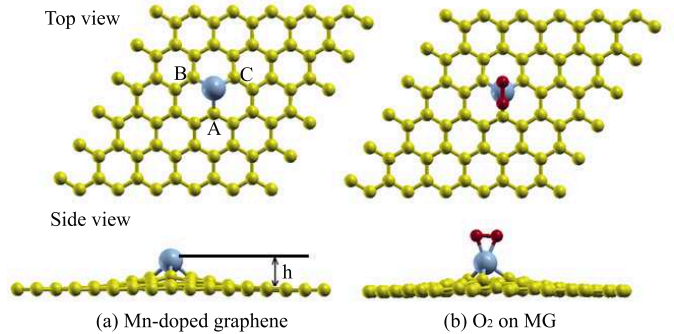}
\caption{(Color online) The most stable configurations of MG with an
elevation $h$ of 1.555 {\AA} (a); and of O$_2$ on MG with $h$ of
1.652 {\AA} (b). Yellow: Carbon, Cambridge blue: Mn, Red: O,
respectively.} \label{config}
\end{figure}

The bond length of O$_2$ molecule in ground state is 1.2350 {\AA}
(experimental value is 1.207 {\AA}) and the magnetic moment is 2
$\mu_B$ in our calculations. O$_2$ molecule is placed at the top of
the doping atoms, and slightly inclined to the graphene plane at the
beginning. After relaxation, O$_2$ molecule is found to be
physisorbed on the top of B and N atoms, while chemisorbed on the
top of Al, Si, P, Cr and Mn atoms. After relaxation, O$_2$ is
parallel to graphene plane, which is the most stable configuration
according to the test of our calculations. The configurations of MG
and O$_2$ adsorption on MG are shown in Fig.\ref{config}, where the
Mn atom protrudes out of the graphene plane. All configurations here
are very similar to Fig.~\ref{config}, except for BG and NG with all
atoms in the same plane and physisorbed O$_2$. In fact, a metastable
configuration with all atoms in one plane can be obtained if the
dopant is set to be in the same plane of graphene, but it is very
easy to transform to the stable one. Meanwhile, the results of O$_2$
adsorption on all doped graphene are shown in Table.\ref{tt}, which
will be discussed detailedly below.

\begin{center}
\begin{table*}[!htb]
\caption{Summary of atomic structures of the compounds, including
adsorption energy $E_a$ (eV), the shortest bond length of X-C
$d_{X-C}$({\AA}), shorter bond length of X-O $d_{X-O}$({\AA}), bond
length of O-O $d_{O-O}$({\AA}), the elevation of the dopant atoms
above the graphene plane $h$ ({\AA}) (negative value means the
dopants protrude out of the plane in opposite direction to O$_2$
molecule), magnetic moments of the system $M_B$ ($\mu_B$), and
L\"{o}wdin charge\cite{Lowdin} transfers from doped graphene to
O$_2$ molecule $c$ ($e^-$). \label{tt}}
\begin{tabular*}{0.99\textwidth}{@{\extracolsep{\fill}}c|cccccccc} \hline\hline
           &           &      Cr &    Mn &  Al   & Si    &    P   &  B     &   N  \\\hline
doped      & $h$       & 1.6487 & 1.5555 & 1.7584& 1.4579& 1.4591 & 0.0000 & 0.0000\\
graphene   & $d_{X-C}$ & 1.8559 & 1.8317 & 1.8528& 1.7611& 1.7689 & 1.4794 & 1.4079\\
           & $M_B$     & 2.0000 & 3.0000 & 0.0000& 0.0000& 1.0500 & 0.0000 & 0.0000\\\hline
O$_2$      & $E_a$     &-2.6098 &-2.0918 &-1.5589&-1.3132&-1.0359 &-0.0232 &-0.1228\\
adsorption & $h$       & 1.6188 & 1.6520 & 1.9026& 1.6607& 1.4840 &-0.2457 &-0.0541\\
           & $d_{X-C}$ & 1.8159 & 1.8464 & 1.9284& 1.8284& 1.7501 & 1.4810 & 1.4066\\
           & $d_{X-O}$ & 1.7935 & 1.8641 & 1.8770& 1.7109& 1.6275 & 3.5099 & 3.3196\\
           & $d_{O-O}$ & 1.4088 & 1.3993 & 1.4000& 1.5103& 1.5584 & 1.2354 & 1.2577\\
           & $M_B$     & AFM    & 1.4500 & 1.0000& 0.3300& 0.0000 & 1.9900 & 1.8000\\
           & $c$       & 0.2166 & 0.1865 & 0.4367& 0.9184& 0.8181 &-0.0863 & 0.0235\\
\hline\hline
\end{tabular*}
\end{table*}
\end{center}

\subsection{adsorption of oxygen on CG and MG}

We first discuss the adsorption of O$_2$ on CG and MG. The atomic
and magnetic structures of CG and MG have been discussed in
Ref.~\onlinecite{rev-tran}. As shown in Fig.~\ref{config} and
Table.~\ref{tt}, the elevation of Cr and Mn atoms is 1.6487 and
1.5555 {\AA}; the Cr-C and Mn-C bond lengths are 1.8559 and 1.8317
{\AA}; the magnetic moments are 2.00$\mu_B$ and 3.00$\mu_B$,
respectively. These are in good agreement with the results using a
larger supercell in Ref.~\onlinecite{rev-tran}. The unpaired
electrons in transition metals (Cr and Mn) induce magnetic moments,
and the direction of spin polarization of Cr or Mn is opposite to
that of the nearest carbon atoms, that is, $C_A$, $C_B$ and $C_C$ in
Fig.~\ref{config}a.

\begin{figure}[!b]
\centering
\includegraphics*[width=3.5in]{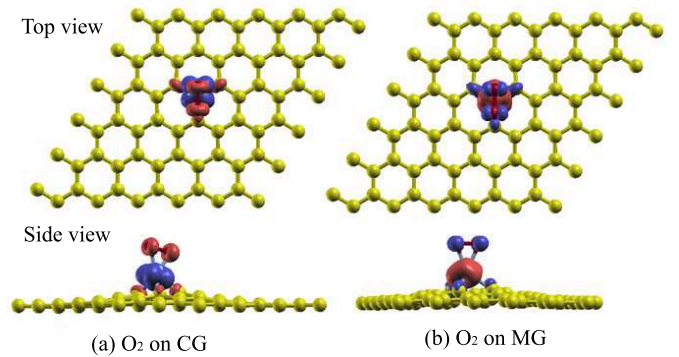}
\caption{(Color online) Spin density in O$_2$ adsorption on doped
graphene. (a) in O$_2$-CG chemisorption system with isovalues of
$\pm$0.01 a.u.; (b) O$_2$-MG chemisorption system with isovalues of
$\pm$0.001 a.u.. Blue color represents spin-down states, red color
spin-up states.} \label{sp-CM}
\end{figure}

From Table.~\ref{tt}, we can learn the configurations are changed
dramatically after O$_2$ adsorption. The molecular oxygen is
chemisorbed on CG and MG with large adsorption energies of -2.6098
and -2.0918 eV, respectively, indicating very stable adsorption of
O$_2$. The bonds of Cr-O and Mn-O are formed with short lengths of
1.7935 and 1.8641 {\AA}; The elevation of Cr above graphene plane is
shortened to 1.6188 {\AA}, while the elevation of Mn is extended to
1.6520 {\AA}. The O-O bond extends to 1.4088 and 1.3993 {\AA} for CG
and MG. Interestingly, CG become AFM by the adsorption of O$_2$,
i.e., with zero total magnetic moment but 0.34$\mu_B$ absolute
magnetic moment. We calculate the magnetic moment of every atom in
the system, and find the magnetic moment of Cr atom (-0.1435) is
opposite to other atoms such as O (0.0411$\mu_B$, 0.0421$\mu_B$),
the three nearest C atoms $C_A$ (0.0133$\mu_B$), $C_B$
(0.0169$\mu_B$) and $C_C$ (0.0169$\mu_B$). At the same time, the
absolute magnetic moment of Cr and the sum of magnetic moment of O,
$C_A$, $C_B$ and $C_C$ is very close. These two properties make the
system AFM, and the magnetization very localized, as shown in
Fig.~\ref{sp-CM}a. For MG, O$_2$ adsorption reduces the magnetic
moment of the system to 1.45 $\mu_B$, which is caused by the
decrease of unpaired electrons due to the hybridization between Mn
and O atoms. Similar to O$_2$ on CG, the magnetic moment is located
around the Mn atom, as shown in Fig.~\ref{sp-CM}b, which may be a
Kondo system.

\begin{figure}[!tb]
\centering
\includegraphics*[width=3.5in]{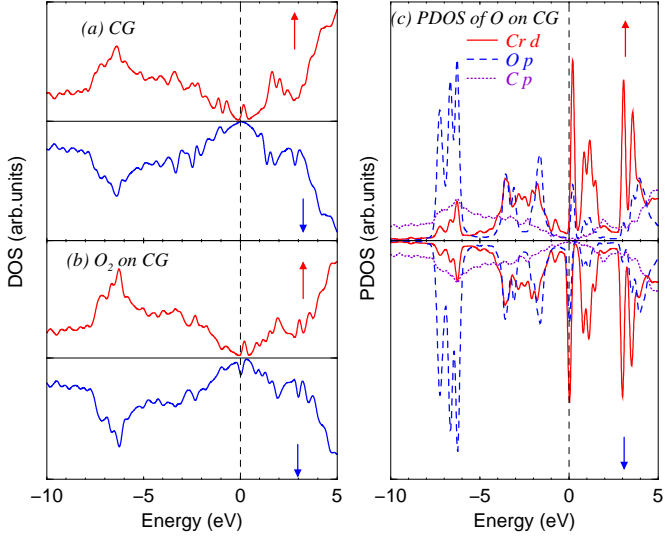}
\caption{(Color online) Total and partial DOS for CG and O$_2$-CG
chemisorption system. (a) Total DOS of CG before adsorption of
O$_2$; (b) Total DOS of CG after adsorption of O$_2$; (c) PDOS of
atoms in O$_2$-CG chemisorbed system. PDOS of O$p$ is the average of
two O atoms, and C$_p$ is the average of all C atoms in the system.
The arrows denote the spin-down ($\downarrow$) states and spin-up
($\uparrow$) states. The zero energy is the Fermi level.} \label{Cr}
\end{figure}

\begin{figure}[!b]
\centering
\includegraphics*[width=3.5in]{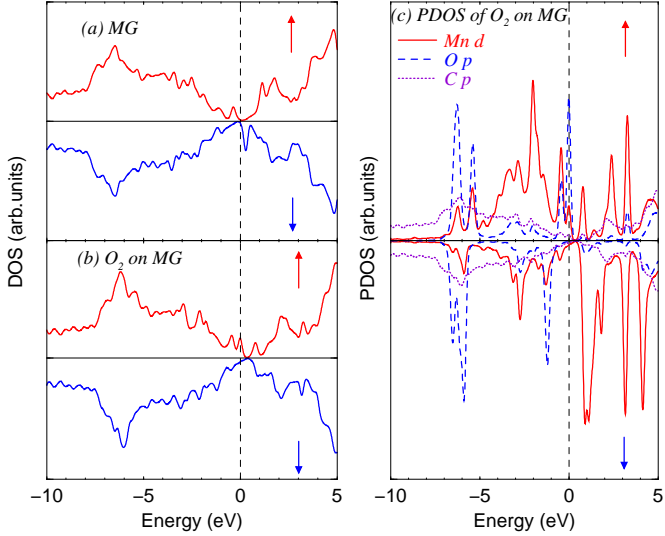}
\caption{(Color online) Total and partial DOS for MG and O$_2$-MG
chemisorption system. (a) Total DOS of MG before adsorption of
O$_2$; (b) Total DOS of MG after adsorption of O$_2$; (c) PDOS of
atoms in O$_2$-MG chemisorbed system. The arrows denote the
spin-down ($\downarrow$) states and spin-up ($\uparrow$) states. The
zero energy is the Fermi level.} \label{Mn}
\end{figure}

In order to understand the electronic properties of the system, DOS
and PDOS of O$_2$-CG and O$_2$-MG are calculated, as shown in
Fig.~\ref{Cr} and ~\ref{Mn}, respectively. It is well known that
intrinsic graphene is semimetal. In Fig.~\ref{Cr}, it is obvious
that CG is a metal. After adsorption of O$_2$, the O, C and Cr
hybridize strongly. There is one DOS peak for spin down electrons
appearing at the Fermi level, which comes from the hybridization of
Cr atom and O$_2$ molecule. One peak disappears below the position
of -1.5 eV, which mainly belongs to the contribution of 3$d$
electrons in Cr atom. L\"{o}wdin charge analysis shows there is
0.2166 $e^-$ transfer from CG to O$_2$ molecule, that is to say,
O$_2$ is an acceptor here. Therefore, O$_2$ can actually p-dope the
host CG, which is consistent with the effect of O$_2$ on organic
semiconductors\cite{hole}. Similarly, O$_2$ can p-dope MG through
accepting 0.1865 $e^-$. From Fig.~\ref{Mn}, we can learn that the
change of electronic properties of MG induced by O$_2$ adsorption is
mainly around the Fermi level, where the DOS peak of spin-down
electrons of MG before adsorption above the Fermi level disappears,
which is contributed by the 3$d$ electrons in Mn atom hybridizing
with 2$p$ electrons in O atom. Meantime, one peak for spin-up
electrons appears at the Fermi level, which is made up of hybridized
Mn$d$ and O$p$ orbitals according to the PDOS analysis in
Fig.~\ref{Mn}. Similar to the analysis of spin density
(Fig.~\ref{sp-CM}), the magnetization is also localized and then
O$_2$ adsorbed on Cr or Mn may exhibit as Kondo impurities as well
as Cr or Mn doped into graphene\cite{rev-tran} and Ni impurity in a
Au nanowire\cite{Kondo}.

\subsection{adsorption of oxygen on AG, SG and PG}

Al, Si and P elements contain 3$p$ orbital, and they have 3, 4 and 5
valence electrons respectively. Discussion about their doping into
graphene and nanotubes has been reported recently, which show that
they are potential resources for detecting toxic molecules and
modulating the electronic structures of graphene.

\begin{figure}[!tb]
\centering
\includegraphics*[width=3.5in]{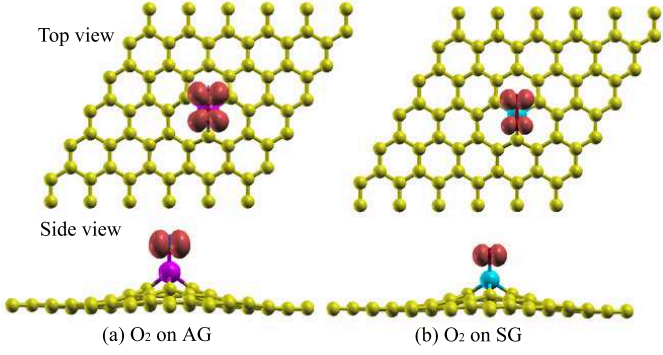}
\caption{(Color online) Spin density in O$_2$ adsorption on doped
graphene. (a) in O$_2$-AG chemisorption system; (b) O$_2$-SG
chemisorption system. Blue color represents spin-down states, red
color spin-up states. The isovalues are $\pm$0.002 a.u. }
\label{sp-AS}
\end{figure}

From Table.~\ref{tt}, we can learn that Al, Si and P atoms introduce
a local curvature of graphene, with elevation of 1.7584, 1.4579 and
1.4591 {\AA}, respectively. Meanwhile, the bond lengths of Al-C,
Si-C and P-C is 1.8528, 1.7611 and 1.7689 {\AA}, respectively, which
basically decrease with the increase of electrons of the elements.
In fact, for atoms which have the same configurations of orbitals,
the covalent bond length decreases with the increase of group
number. This indicates that the extension of electron states plays a
very important role in the structures of doped graphene. Specially,
P dopant introduces spin-polarization into the PG with magnetic
moment of 1.05 $\mu_B$, while no magnetization exists for AG and SG.
It is noticed that a metastable configuration of PG with all atoms
retaining in one plane can be observed, in which there is no spin
polarization induced. It seems that the local curvature has a very
important effect for the magnetic property. O$_2$ molecule can be
chemisorbed on the atoms of Al, Si and P in graphene, with large
adsorption energies of -1.5589, -1.3132 and -1.0359 eV,
respectively. Meantime, the atomic structures change much. For AG,
Al atom has an elevation of 1.9026 {\AA}, and C-Al bond extends to
1.9284 {\AA}, Al-O to 1.8770 {\AA} and O-O to 1.40 {\AA}. For SG, Si
atom protrudes out of graphene plane with elevation of 1.6607 {\AA}.
C-Si, Si-O and O-O bonds also elongate, which are 1.8284, 1.7109 and
1.5103 {\AA}, respectively. On the contrary, chemisorption of O$_2$
on PG shorten the P-C bond to 1.6286 {\AA} from 1.7691 {\AA} before
adsorption, and the bond O-O is broken completely with the long
distance of 1.5598 {\AA}. The P atom almost retains the place above
graphene with 1.4796 {\AA} and does not protrude outward more. It is
interesting that chemisorption of O$_2$ introduces spin polarization
for AG and SG with magnetic moments of 1.00 and 0.33 $\mu_B$, but
the magnetization in PG vanishes after O$_2$ adsorption. For AG and
SG, the spin-polarization is mainly located in the O atoms, which is
also very localized as shown in Fig.~\ref{sp-AS}. Therefore, O$_2$
adsorption may introduce Kondo effect into AG and SG systems.

\begin{figure}[!tb]
\centering
\includegraphics*[width=3.5in]{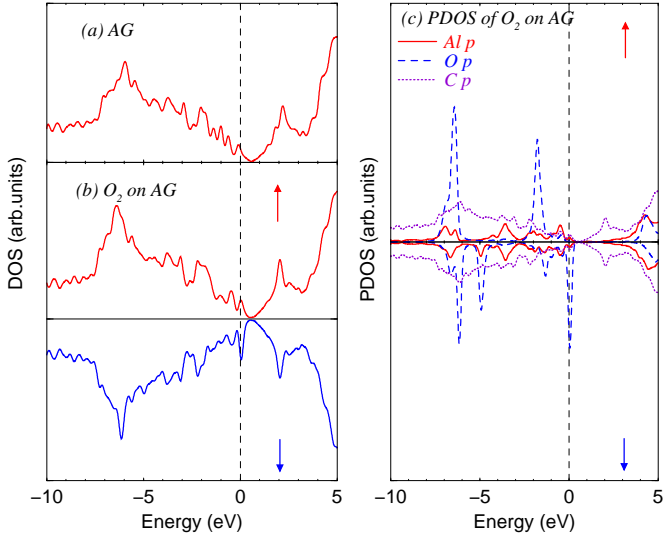}
\caption{(Color online) Total and partial DOS for AG and O$_2$-AG
chemisorption system. (a) Total DOS of AG before adsorption of
O$_2$; (b) Total DOS of AG after adsorption of O$_2$; (c) PDOS of
atoms in chemisorbed system. The arrows denote the spin-down
($\downarrow$) states and spin-up ($\uparrow$) states. The zero
energy is the Fermi level.} \label{Al}
\end{figure}

The electronic properties are also investigated by the analysis of
DOS (PDOS) and L\"{o}wdin charges. For AG after adsorption, the
analysis of L\"{o}wdin charges shows that O$_2$ molecule accepts
about 0.44 $e^-$ from Al atom, and 0.72 $e^-$ transfers from Al to
the three nearest C atoms (0.24 per atom). The DOS of AG before and
after adsorption of O$_2$ is shown in Fig.~\ref{Al}. Al has 3
valence electrons, and therefore can be hole-doping for graphene, as
shown in Fig.~\ref{Al}a, where the minimum of DOS shifts above the
Fermi level. Chemisorption of O$_2$ introduces spin polarization,
which creates the difference of the DOS mainly around the Fermi
level, as shown in Fig.~\ref{Al}b. The peak at Fermi level for spin
down electrons is contributed by the O$p$ orbital according to the
PDOS analysis, shown in Fig.~\ref{Al}c.

\begin{figure}[!tb]
\centering
\includegraphics*[width=3.5in]{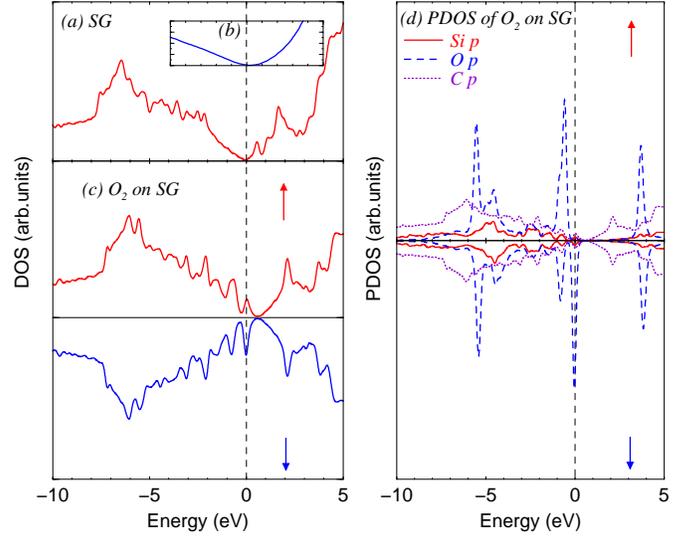}
\caption{(Color online) Total and partial DOS for SG and O$_2$-SG
chemisorption system. (a) Total DOS of SG before adsorption of
O$_2$; (b) DOS of SG with 3\% concentration; (c) Total DOS of SG
after adsorption of O$_2$; (d) PDOS of atoms in chemisorbed system.
The arrows denote the spin-down ($\downarrow$) states and spin-up
($\uparrow$) states. The zero energy is the Fermi level.} \label{Si}
\end{figure}

For chemisorption of O$_2$ on SG, O$_2$ molecule accepts about 0.92
$e^-$ from Si atom, and about 0.23 $e^-$ transfers from Si atom to
every nearest C atom. Si has 4 electrons, the same as C atom.
Therefore, the doping of Si should not shift the Fermi level of
graphene, as shown in Fig.~\ref{Si}a. It is worth pointing out that
the electronic structures are dependent on the dopant concentration,
as shown in Fig.~\ref{Si}b, and the electronic band of SG system is
open with a gap of 0.1 eV. Nevertheless, it does not affect the
configuration of chemisorption of O$_2$ on SG. O$_2$ adsorption
introduces magnetization with 0.33 $\mu_B$. As shown in
Fig.~\ref{Si}c-d, the difference of the DOS between spin up and spin
down electrons is mainly caused by O$p$ orbital. The DOS peak of
spin up electrons at the Fermi level is caused by the $p_z$ orbital
in C atom, while the peak of spin down by O$p$. Moreover, the
hybridization of Si$p$, O$p$ and C$p$ orbitals also happens here.

\begin{figure}[!tb]
\centering
\includegraphics*[width=3.0in]{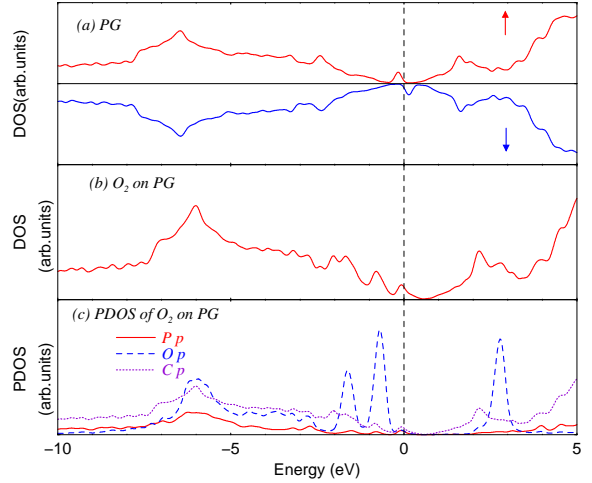}
\caption{(Color online) Total and partial DOS for PG and O$_2$-PG
chemisorption system. (a) Total DOS of PG before adsorption of
O$_2$; (b) Total DOS of PG after adsorption of O$_2$; (c) PDOS of
atoms in chemisorbed system. The arrows denote the spin-down
($\downarrow$) states and spin-up ($\uparrow$) states. The zero
energy is the Fermi level.} \label{Ph}
\end{figure}
\par

For O$_2$ on PG, about 0.82 $e^-$ transfers from P atom to two O
atoms, and about 0.24 $e^-$ from P atom to every nearest C atom.
Before O$_2$ binding, PG is magnetic with 1.05 $\mu_B$. Here, charge
analysis shows us that every nearest C atom accepts about 0.18 $e^-$
from P atoms, and about 4.3 $e^-$ remains in P atoms. That is to
say, there is about one electron unpaired, inducing about 1 $\mu_B$
magnetic moment. After adsorption of O$_2$, the hybridization
between P and O, even inluding C, employs the unpaired electron, and
then the spin polarization vanishes. According to the DOS of PG
before adsorption of $O_2$, as shown in Fig.~\ref{Ph}a, the
difference of DOS between spin up and spin down electrons mainly
locates around the Fermi level. The binding of O$_2$ evaporates the
magnetization and creates a peak at the position of -0.75 eV, as
shown in Fig.~\ref{Ph}b, which is caused by the hybridization of
P$p$, O$p$ and C$p$ orbitals, as shown in Fig.~\ref{Ph}c.

\subsection{Adsorption of oxygen on BG and NG}

\begin{figure}[!tb]
\centering
\includegraphics*[width=3.2in]{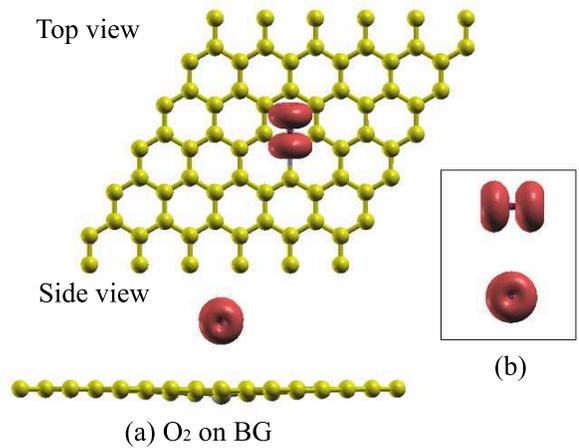}
\caption{(Color online) Spin density in O$_2$ adsorption in (a)
O$_2$-BG physisorption system, and (b) isolated O$_2$. The isovalues
are $\pm$0.005 a.u. } \label{sp-bn}
\end{figure}

B and N belong to the second group elements in the periodic table,
which have $2s2p$ configurations for electrons and are the nearest
elements for C. The extension of electrons of B, C and N is similar.
Therefore, there is no local distortion when B and N are doped in
graphene. The length of B-C is 1.4794 {\AA}, larger than that of N-C
with 1.4079 {\AA}. B and N atoms do not have elevation above
graphene, and there is also no spin polarization. O$_2$ is
physisorbed on BG and NG, which is the same as O$_2$ on B-doped
graphite\cite{b-o}. The physisorption does not change the structures
of BG and NG significantly, and the magnetic moments of O$_2$ is
also almost retained, as shown in Fig.~\ref{sp-bn}a. The spin
density of O$_2$-NG is very similar. The spin density of O$_2$-BG
system concentrates in the O$_2$ molecule, which is almost the same
as the spin density of isolated O$_2$ molecule, shown in
Fig.~\ref{sp-bn}b. The O$_2$ molecule is far away from graphene with
a distance larger than 3.5 {\AA} and a relatively small adsorption
energy. Furthermore, B and N atoms is slightly pushed below graphene
plane by O$_2$, indicating the repulsed force between them. Since B
and Al, N and P belong to the same group elements and have very
similar valence electron configurations, and their behaviors caused
by O$_2$ molecule are so different, it can be induced that one
important reason for this phenomenon is the local curvature in AG
and PG. This is also in agreement with the conclusion of local
curvature enhancing chemical reactivity of doped
graphene\cite{Garcia}. About this local curvature, as in
Table.~\ref{tt}, only the B and N dopants are not elevated, and the
B-C and N-C bond lengths are close to C-C bond in graphene. By
contrary, other dopants are elevated, and the X-C bonds are much
longer than C-C bonds. Therefore, it can be concluded that dopant X
is elevated only when the length of X-C is appreciably larger than
that of C-C.

\begin{figure}[!tb]
\centering
\includegraphics*[width=3.2in]{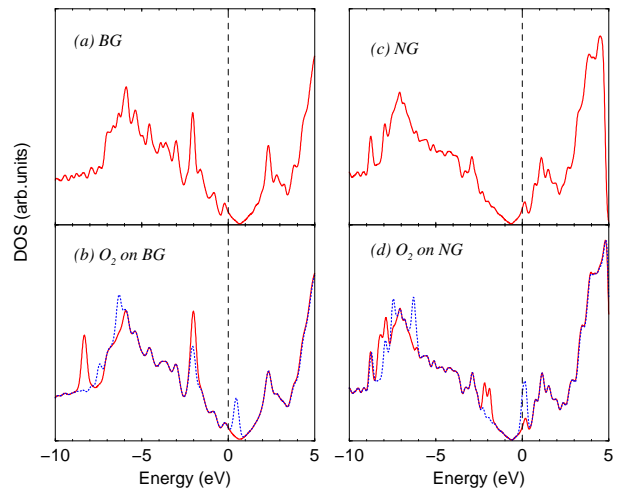}
\caption{(Color online) (a) Total DOS of BG before adsorption of
O$_2$; (b) Total DOS of BG after adsorption of O$_2$; (a) Total DOS
of NG before adsorption of O$_2$; (d) Total DOS of NG after
adsorption of O$_2$; Red solid line represents spin-up states and
blue dashed line spin-down states in (b) and (d). The zero energy is
the Fermi level.} \label{np}
\end{figure}

Although O$_2$ can not bind stably on BG and NG, it can still affect
the electronic properties at low temperature at least. The dopant of
Boron can hole-dope graphene and Nitrogen electron-dope graphene, as
shown in Fig.~\ref{np}a,c, where the Dirac point shifts above (BG)
or below (NG) the Fermi level. After adsorption of $O_2$, the spin
polarization is introduced and the DOS is changed. One DOS peak
around Fermi level appears for spin down electrons of BG and NG
physisorbed O$_2$ molecule, caused by O$p$ orbital; but the DOS of
spin up electrons is very similar to that of BG before adsorption.
This means that the change in DOS is concentrated on the O$_2$
molecule. However, this effect should not exist at room temperature
because of the very weak adsorption and the long distance between
O$_2$ and graphene. Furthermore, the charge transfer between O$_2$
and BG or NG is very small compared with the chemisorption,
indicating the small change for the conductance of the system. It
should be noticed that O$_2$ acts as a donor for BG through donating
$\sim$0.08 $e^-$ to BG system. Although the charge transfer of the
physisorbed system using the figure of the single electron Kohn-Sham
equation is under debate, the trend of this small charge transfer is
reliable.

\section{Conclusion}

In conclusion, our simulation studies reveal that most of doped
graphene are sensitive to O$_2$ molecule. The local curvature caused
by doped atoms is very important to the reactivity for gas
molecules. BG and NG are inert to molecular oxygen, while graphene
with Al, Si, P, Cr and Mn doping allows O$_2$ to form stable
chemisorption states, which affects the magnetic, electronic and
atomic properties of graphene. O$_2$ acts as an acceptor in all
chemisorbed configurations, and introduces localized spin
polarization except on PG. In particular, O$_2$ on CG is AFM, and
O$_2$ can retain its magnetic moments on B- and N-doped graphene.
Significantly, O$_2$ adsorption may introduce Kondo effect into CG,
MG, AG and SG systems. Most of systems with O$_2$ on doped graphene
behave as metallic materials. Therefore, a combination of foreign
atom doping followed by exposure to air may be an effective way to
tune the electronic and magnetic properties of semimetal and
unpolarized graphene. However, the potential usage of doped graphene
as gas sensors should be dependent on the sensitivity to O$_2$
molecule in air. When chemisorption of O$_2$ on doped graphene
happens, the sensitivity to other gases should be affected
significantly, which may prevent it from being effective gas
sensors.

\section{Acknowledgments}

This work is supported by the National Natural Science Foundation of
China under Grant Nos. 10734140 and 60621003, the National Basic
Research Program of China (973 Program) under Grant No.
2007CB815105, and the National High-Tech ICF Committee in China. All
calculations are carried at the Research Center of Supercomputing
Application, NUDT.


\begin{thebibliography}{9}
\bibitem{first} K. S. Novoselov, A. K. Geim, S. V. Morozov, D. Jiang, Y.
Zhang, S. V. Dubonos, I. V. Grigorieva, and A. A. Firsov, Science
{\bf 306}, 666 (2004).
\bibitem{Rise} A. K. Geim and K. S. Novoselov, Nat. Mater. {\bf
6}, 183 (2007).
\bibitem{Neto} A. H. Castro Neto, F. Guinea, N. M. R. Peres, K. S. Novoselov and A. K. Geim,
Rev. Mod. Phys. {\bf 81}, 109 (2009).
\bibitem{Schedin} F. Schedin, A. K. Geim, S. V. Morozov, E. W. Hill, P. Blake, M. I. Katsnelson
and K. S. Novoselov, Nat. Mater. {\bf 6}, 652 (2007).
\bibitem{Novoselov} K. S. Novoselov, A. K. Geim, S. V. Morozov, D. Jiang, M. I. Katsnelson, I. V. Grigorieva,
S. V. Dubonos and A. A. Firsov, Nature {\bf 438}, 197 (2005).
\bibitem{Zhang} Y. Zhang, J. Tan, H. L. Stormer and P. Kim, Nature {\bf 438}, 201 (2005).
\bibitem{Danneau} R. Danneau, F. Wu, M. F. Craciun, S. Russo,
M. Y. Tomi, J. Salmilehto, A. F. Morpurgo and P. J. Hakonen, Phys.
Rev. Lett. {\bf 100}, 196802 (2008).
\bibitem{control} T. Ohta, A. Bostwick, T. Seyller, K. Horn and E. Rotenberg, Science {\bf 313}, 951 (2006).
\bibitem{spin} T. G. Pedersen, C. Flindt, J. Pedersen, N. A. Mortensen, A. P.
Jauho, and K. Pedersen, Phys. Rev. Lett. {\bf 100}, 136804 (2008);
B. Trauzettel, D. V. Bulaev, D. Loss, and G. Burkard, Nat. Phys.
{\bf 3}, 192 (2007); A. Rycerz, J. Tworzyd, and C. W. J. Beenakker,
Nat. Phys. {\bf 3}, 172 (2007).
\bibitem{sensor} B. Huang, Z. Li, Z. Liu, G. Zhou, S. Hao, J. Wu, B.-
L. Gu, and W. Duan, J. Phys. Chem. C {\bf 112}, 13442 (2008); R.
Moradian, Y. Mohammadi, and N. Ghobadi, J. Phys. Cond. Matt. {\bf
20}, 425211 (2008).
\bibitem{Leenanerts} O. Leenaerts, B. Partoens and F. M. Peeters, Phys. Rev. B
{\bf 77}, 125416 (2008).
\bibitem{Paolo} P. Giannozzi, R. Car and G. Scoles,
J. Chem. Phys. {118}, 1003 (2003).
\bibitem{OurPaper} J. Dai, P. Giannozzi and J. Yuan, Surf. Sci. {\bf 603}, 3234 (2009).
\bibitem{LDA} Y. Zhang, Y. Chen, K. Zhou, C. Liu, J. Zeng, H. Zhang
and Y. Peng, Nanotechnology {\bf 20}, 185504 (2009).
\bibitem{apl} J. Dai, J. Yuan and P. Giannozzi, Appl. Phys. Lett. {\bf 95}, 232105 (2009).
\bibitem{Rangel} E. Rangel, G. R. Chavarria and L. F. Magana, Carbon {\bf 47}, 531 (2009).
\bibitem{si} H. Jiang, D. Zhang and R. Wang, Nanotechnology {\bf 20},
145501 (2009).
\bibitem{half-B} S. Dutta and S. K. Pati, J. Phys. Chem. B, {\bf 112}, 1333
(2008).
\bibitem{Garcia} A. L. E. Garcia, S. E. Baltazar, A. H. Romero, J. F. Perez Robles
and A. Rubio, J. Comput. Theor. Nanosci. {\bf 5}, 2221 (2008).
\bibitem{tran} N. Gorjizadeh, A. A. Farajian, K. Esfarjani and Y.
Kawazoe, Phys. Rev. B {\bf 78}, 155427 (2008); E. J. G. Santos, A.
Ayuela, S. B. Fagan, J. Mendes Filho, D. L. Azevedo, A. G. Souza
Filho and D. S\'{a}nchez-Portal, Phys. Rev. B {\bf 78}, 195420
(2008).
\bibitem{rev-tran} A. V. Krasheninnikov, P. O. Lehtinen, A. S. Foster, P. Pyykko and R. M.
Nieminen, Phys. Rev. Lett. {\bf 102}, 126807 (2009).
\bibitem{zhong} Y. Mao and J. Zhong, Nanotechnology {\bf 19}, 205708
(2008).
\bibitem{defect} S. Lakshmi, S. Roche, and G. Cuniberti, Phys. Rev. B {\bf 80}, 193404 (2009).
\bibitem{func} A. Nduwimana and X. Q. Wang, ACS Nano {\bf 3}, 1995 (2009); G. Cantele, Y. S.
Lee, D. Ninno, and N. Marzari, Nano Lett. {\bf 9}, 3425 (2009); Y.
H. Zhang, K. G. Zhou, K. F. Xie, J. Zeng, H. L. Zhang, and Y. Peng,
Nanotechnology {\bf 21}, 065201 (2010).
\bibitem{Nitr} D. Wei, Y. Liu, Y. Wang, H. Zhang, L. Huang and G.
Yu, Nano. Lett. {\bf 9}, 1752 (2009).
\bibitem{bn} L. S. Panchakarla, K. S. Subrahmanyam, S. K. Saha, A. Govindaraj, H. R.
Krishnamurthy, U. V. Waghmare and C. N. R. Rao, Adv. Mater. {\bf
21}, 4726 (2009).
\bibitem{pho} I. O. Maciel, J. Campos-Delgado, E. Cruz-Silva, M. A. Pimenta, B. G.
Sumpter and V. Meunier, F. L\'{o}pez-Ur\'{\i}as, E.
Mu\~{n}oz-Sandoval, H. Terrones, M. Terrones and A. Jorio, Nano.
Lett. {\bf 9}, 2267 (2009).
\bibitem{hole} C. Lu and H. Meng, Phys. Rev. B {\bf 75}, 235206
(2007).
\bibitem{gas} J. Berashevich and T. Chakraborty, Phys. Rev. B {\bf
80}, 033404 (2009); T.O. Wehling et al, Nano Lett. {\bf 8}, 173
(2008); S.S. Yu et al, IEEE TRANSACTIONS ON NANOTECHNOLOGY {\bf 7},
628 (2008).
\bibitem{exp} H. Ulbricht, G. Moos and T. Hertel, Surf. Sci. {\bf
532}, 852 (2003).
\bibitem{cbn} J. Zhang, K. P. Loh, J. Zheng, M. B. Sullivan and P.
Wu, Phys. Rev. B {\bf 75}, 245301 {2007}.
\bibitem{PBE} J. P. Perdew, K. Burke and M. Ernzerhof, Phys. Rev. Lett. {\bf 77}, 3865 (1996).
\bibitem{single} P. M. Voyles, D. A. Muller, J. L. Grazul, P. H.
Citrin, and H. J. L. Gossmann, Nature {\bf 416}, 826 (2002); C. I.
Pakes, D. P. George, D. N. Jamieson, C. J. Yang, A. S. Dzurak, E.
Gauja and R. G. Clark, Nanotechnology {\bf 14}, 157 (2003).
\bibitem{pot} The pseudopotentials of B.pbe-n-van.UPF, C.pbe-van$_{-}$bm.UPF, O.pbe-van$_{-}$bm.UPF, N.pbe-van$_{-}$bm.UPF,
Al.pbe-n-van.UPF, Si.pbe-n-van.UPF, P.pbe-n-van.UPF,
Cr.pbe-sp-van.UPF and Mn.pbe-sp-van.UPF from the Quantum ESPRESSO distribution.
\bibitem{MK} H. J. Monkhorst and J. D. Pack, Phys. Rev. B {\bf 13}, 5188 (1976).
\bibitem{MP} M. Methfessel and A. T. Paxton, Phys. Rev. B {\bf 40}, 3616 (1989).
\bibitem{tetrahedron} P. E. Bl\"{o}chl, O. Jepsen and O. K.
Andersen, Phys. Rev. B {\bf 49}, 16223 (1994).
\bibitem{pwscf} P. Giannozzi \textit{et al}, J. Phys.: Condens. Matter
{\bf 21} 395502 (2009). URL: \url{www.quantum-espresso.org}.
\bibitem{Lowdin} P. O. L\"{o}wdin, J. Chem. Phys. {\bf 18}, 365
(1950).
\bibitem{Kondo} P. Lucignano, R. Mazzarello, A. Smogunov, M.
Fabrizio and E. Tosatti, Nat. Mater. {\bf 8}, 563 (2009); K.
Sengupta and G. Baskaran, Phys. Rev. B {\bf 77}, 045417 (2008).
\bibitem{b-o} Y. Ferro, A. Allouche, F. Marinelli and C.
Brosset, Surf. Sci. {\bf 559}, 158 (2004).
\end{thebibliography}
\end{document}